# A Music-generating System Inspired by the Science of Complex Adaptive Systems

**Shawn Bell**
A.L.C Program, Interactive Media Arts
Dawson College, Canada

**Liane Gabora**
Psychology Department
University of British Columbia , Canada

## Abstract

This paper presents NetWorks (NW), an interactive music-generation system that uses a hierarchically clustered scale-free network to generate music that ranges from orderly to chaotic. NW was inspired by the Honing Theory of creativity, according to which human-like creativity hinges on (1) the ability to self-organize and maintain dynamics at the 'edge of chaos' using something akin to 'psychological entropy', and (2) the capacity to shift between analytic and associative processing modes. At the 'edge of chaos' NW generates patterns that exhibit emergent complexity through coherent development at low, mid, and high levels of musical organization, and often suggests goal seeking behavior. The architecture consists of four 16-node modules: one each for pitch, velocity, duration, and entry delay. The Core allows users to define how nodes are connected, and rules that determine when and how nodes respond to their inputs. The Mapping Layer allows users to map node output values to MIDI data that is routed to software instruments in a digital audio workstation. By shifting between bottom-up and top-down NW shifts between analytic and associative processing modes.

## Introduction

This paper introduces NetWorks (NW), a music-generating program inspired by the view that (1) the human mind is a complex adaptive system (CAS), and thus (2) human-like computational creativity can be achieved by drawing on the science of complex systems. NW uses scale-free networks and 'edge of chaos' dynamics to generate music that is aesthetically pleasing and maintains interest. The approach dates back to a CD of emergent, self–organizing computer music based on cellular automata and asynchronous genetic networks titled "Voices From The Edge of Chaos" (Bell 1998), and more generally to the application of artificial life models to computer-assisted composition, generative music, and sound synthesis (Beyls 1989, 1990, 1991; Bowcott 1989; Chareyron 1990; Horner and Goldberg 1991; Horowitz 1994; Millen 1992; Miranda 1995; Todd and Loy 1991).

We first summarize key elements of a CAS-inspired theory of creativity, and discuss the relevance for computational creativity. Next we outline the architecture of NW, evaluate its outputs, and highlight some of its achievements. We then summarize how NW adheres to principles of honing theory and CAS, and how this contributes to the appealing musicality of its output.

## The Honing Theory of Creativity

The honing theory (HT) of creativity (Gabora 2010, in press) has its roots in the question of what kind of structure could evolve novel, creative forms effectively and strategically (as opposed to at random). We now summarize the elements of the theory most relevant to NetWorks.

### Mind as a Self-Organizing Structure

Humans possess two levels of complex, adaptive, structure: an organismic level and a psychological level, i.e., a mind (Pribram 1994). Like a body, a mind is self-organizing: a new stable global organization can emerge through interactions amongst the parts (Ashby 1947; Carver and Scheier 2002; Prigogine and Nicolis 1977). The capacity to self-organize into a new patterned structure of relationships is critical for the generation of creative outcomes (Abraham 1996; Goertzel 1997; Guastello 1998). The mind's self-organizing capacity originates in a memory that is distributed, content addressable, and sufficiently densely packed that for any one item there is a reasonable probability it is similar enough to some other item to evoke a reminding of it, thereby enabling the redescription and refinement of ideas and actions in a stream of thought (Gabora, 2000, 2010). Mental representations are distributed across neural cell assemblies that encode for primitive stimulus features such as particular tones or tim-bres. Mental representations are both constrained and enabled by the connections between neurons they activate.

Just as a body mends itself when injured, a mind is on the lookout for 'gaps'—arenas of incompletion or inconsistency or pent-up emotion—and explores the gap from different perspectives until a new understanding has been achieved. We use the term *self-mending* to refer to the capacity to reduce psychological entropy in response to a perturbation (Gabora, in press), i.e., it is a form of self-organization involving reprocessing of arousal-provoking material. Creative thinking induces restructuring of representations, which may involve re-encoding the task such that new elements are perceived to be relevant, or relaxing goal constraints. The transformative impact of immersion in a creative process can bring about sweeping changes to the second (psychological) level of complex, adaptive structure, that alter one's self-concept and worldview.

## The Edge of Chaos

Self-organized criticality (SOC) is a phenomenon wherein, through simple local interactions, complex systems find a critical state poised at the transition between order and chaos—the proverbial edge of chaos—from which a small perturbation can exert a disproportionately large effect (Bak, Tang, and Weisenfeld 1988). It has been suggested that insight is a self-organized critical event (Gabora 1998; Schilling 2005). SOC gives rise to structure that exhibits sparse connectivity, short average path lengths, strong local clustering, long-range correlations in space and time, and rapid reconfiguration in response to external inputs. There is evidence of SOC in the human brain, e.g., with respect to phase synchronization of large-scale functional networks (Kitbiczler, Smith, Christensen, and Bullmore 2009). There is also evidence of SOC at the cognitive level; word association studies show that concepts are clustered and sparsely connected, with some having many associates and others few (Nelson, McEvoy, and Schreiber 2004). Cognitive networks exhibit the sparse connectivity, short average path lengths, and local clustering characteristic of self-organized complexity and in particular 'small world' structure (Steyvers and Tenenbaum 2005).

Like other SOC systems, a creative mind may function within a regime midway between order (systematic progression of thoughts), and chaos (everything reminds one of everything else). Much as most perturbations in SOC systems have little effect but the occasional perturbation has a dramatic effect, most thoughts have little effect on one's worldview, but occasionally one thought triggers another, which triggers another, and so forth in a chain reaction of conceptual change. This is consistent with findings that large-scale creative conceptual change often follows a series of small conceptual changes (Ward, Smith, and Vaid 1997), and with evidence that power laws and catastrophe models are applicable to the diffusion of innovations (Jacobsen and Guastello 2011).

## Two Modes of Thought: Contextual Focus

Psychological theories of creativity typically involve a divergent stage that predominates during idea generation and a convergent stage that predominates during the refinement, implementation, and testing of an idea (for a review see Runco 2010; for comparison between divergent / convergent creative processes and dual process models of cognition see Sowden, Pringle, and Gabora 2015). Divergent thought is characterized as intuitive and reflective; it involves the generation of multiple discrete, often unconventional possibilities. It is contrasted with convergent thought, which is critical and evaluative; it involves tweaking of the most promising possibilities. There is empirical evidence for oscillations in convergent and divergent thinking, with a relationship between divergent thinking and chaos (Guastello 1998). It is widely believed that divergent thought involves defocused attention and associative processing, and this is consistent with the literal meaning of divergent as "spreading out" (as in a divergence of a beam of light). However, the term divergent thinking has come to refer to the kind of thought that occurs during creative tasks that involve the generation of multiple solutions, which may or may not involve defocused attention and associative memory. Moreover, in divergent thought, the associative horizons simply widen generically instead of in a way that is tailored to the situation or context (Fig. 2). Therefore, we will use the term associative thought to refer to creative thinking that involves defocused attention and context-sensitive associative processes, and analytic thought to refer to creative thinking that involves focused attention and executive processes. The capacity to shift between these modes of thought has been referred to as contextual focus (CF) (Gabora 2010). While some dual processing theories (e.g., Evans 2003) make the split between automatic and deliberate processes, CF makes the split between an associative mode conducive to detecting relationships of correlation and an analytic mode conducive to detecting relationships of causation. Defocusing attention facilitates associative thought by diffusely activating a broad region of memory, enabling obscure (though potentially relevant) aspects of a situation to come to mind. Focusing attention facilitates analytic thought by constraining activation such that items are considered in a compact form amenable to complex mental operations.

According to HT, because of the architecture of associative memory, creativity involves not searching and selecting amongst well-formed idea candidates, but amalgamating and honing initially ill-formed possibilities from multiple sources. As a creative idea is honed, its representation changes through interaction with internally or externally generated contexts, until psychological entropy is acceptably low. The unborn idea is said to be in a 'state of potentiality' because it could actualize different ways depending on the contextual cues taken into account as it takes shape.

## The NetWorks Musical System

NW consists of a music-generating system and the music it has produced. The goals of NW are to (1) generate "emergent music," i.e., self-organizing, emergent dynamics from simple rules of interaction, expressed in musical forms, and (2) through emergence, discover new genres of music. In terms of creative agency, NW has been designed as a closed, autonomous system while generating MIDI data. In selecting the network architecture and interaction rules, the artist-user may be viewed as the system's mentor. The MIDI data generated by the system is orchestrated and mixed by the artist-user, who may be viewed in this role as a collaborator (McCormack and d'Inverno 2014).

Network theory, as it pertains to the study of complex adaptive systems (Mitchell 2006) was used in the design of the NW system. NW is currently configured to explore the expressive potential of hierarchical scale-free networks, as the properties of such networks underlies the interesting dynamics of many real world networks, from the cell to the World Wide Web (Barabási 2002). Moreover, musical genres that appear to be very different on the surface have been shown to exhibit an underlying scale-free structure. Music composed by Bach, Chopin and Mozart, as well as Russian folk and Chinese pop music have been shown to be scale-free (Liu, Tse & Small 2009). Given the ubiquity of hierarchical scale-free topology and dynamics found in CAS it is not surprising that such architectures have creative potential.

NW is composed of two layers (1) the Core, which allows the artist-user to define how the nodes are connected, as well as the rules that determine when and how nodes respond to their inputs, and (2) the Mapping Layer, which allows the artist-user to map node output values to MIDI data that are routed to software instruments in a Digital Audio Workstation (DAW).

We now discuss these two layers in more detail.

### The Core

A note has five basic attributes: pitch, loudness (usually corresponding to MIDI velocity), duration, timing (or entry delay), and timbre. The core consists of 64 nodes connected in a scale-free architecture, organized into four 16-node modules: one for pitch, velocity, duration, and entry delay (Figure 1).

Pitch nodes output values for pitch, but require values for velocity and duration to produce a note. The nodes of the velocity and duration modules provide these values. Four, sixteen node modules allow for 16 channels of MIDI output and the timbral characteristics for each stream of notes is determined by the artist-user by their choice of instruments.

The entry delay (ED) module is responsible for keeping the corresponding nodes of the four modules synchronized (Figure 1). When a pitch node is activated, as determined by the delay value it receives from its ED module node, the corresponding velocity and duration module nodes are activated simultaneously to provide the values required to specify a note. The function of the ED module is to determine timing, that is, when nodes produce an output, and therefore the pattern of activation across the network as a whole. In musical terms, the entry delay module generates rhythmic patterns via note groupings, from motivic cells to entire movements.

When the nodes are fully connected—that is, receiving values on all their inputs—the network architecture is scale-free; however users can prune the connectivity of the network by reducing the number of inputs to the nodes.

**Nodes**

The values that nodes send and receive are integers, within a range specified by the artist-user, for example 1–13, 1-25, etc. Nodes function according to the following algorithm:

1. nodes store the most recently received values from connected nodes;
2. when a node is activated the values are summed;
3. the sum is sent to a look-up table which outputs a value;
4. the value is delayed, as determined by the corresponding node in the ED module;
5. the value is then sent to connected nodes, as well as back to the originating node.

The pitch module is unique; it includes the largest hub, which sends values to, and receives values from, 40 nodes: 12 pitch nodes, 9 nodes each from the duration, velocity, and ED modules, as well as from itself.

It is important to emphasize that hubs receive values from, and send values to, hubs in other modules. In this way, note attributes affect one another as the music evolves over time, for example: duration can influence pitch, pitch can influence entry delay, entry delay can influence velocity, and so on.

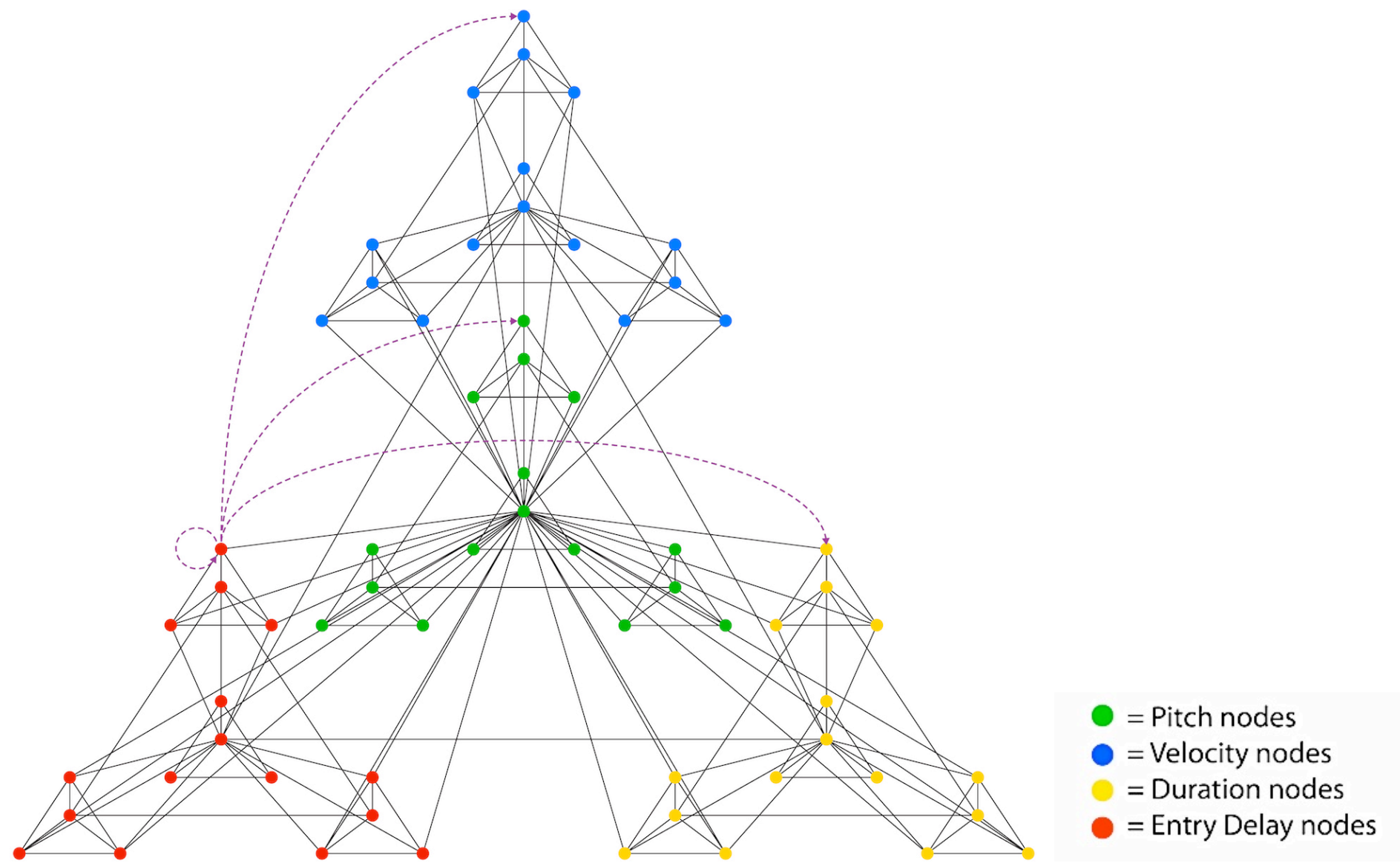

Figure 1. Schematic illustration of the different kinds of nodes and their interrelationships. Undirected edges (in black) indicate that values can be exchanged in both directions, i.e.,nodes both send values to, and receive values from, nodes to which they are connected. Directed edges (purple) show the relationship between individual nodes of the Entry Delay module and the corresponding nodes of other modules. The ED module node determines *when* it will activate itself, and the corresponding node in the duration, velocity, and pitch modules. For clarity, only one of the 16 ED nodes and its four corresponding nodes are shown.

**Rules**

When activated, a node sums the last values it received and sends the sum to a look-up table (LUT) that returns the value stored at the corresponding index. Experiments have focused on the range of integers from 1–25 that allows for pitch to be mapped chromatically across two octaves and provides the same number of equivalent scale steps for velocity, duration and entry delay.

NetWorks has been designed to allow (1) each node to have its own LUT, (2) an LUT for each module, and (3) one LUT for all the nodes of the network. LUTs are generated using a variety of methods: random, random without repetition, ratios, etc.

The dynamics of the network are controlled by the choice of LUTs. Networks with nodes using LUTs with a random distribution of output values result in chaotic MIDI data sequences. Networks with nodes using LUTs that output the same value produce total repetition.

**Relationship between Architecture and Rules**

Two observations can be made regarding the relationship of rules and network architecture. First, when the network is scale-free, nodes have either 4, 5, 6, 15 or 40 inputs. Each input on a node can receive a range of values that are summed to determine an output value (via the LUT). This means the range of output values is always less than the range of possible summed values, which results in a loss of "information." For example, the largest hub with 40 inputs requires an LUT with an index of 520, but can only output 13 different values. Hence, nodes can be thought of as "funnels," in that the range of integers they can output is always less than the range of possible sums of their inputs.

Second, while hubs have a wider "sphere of influence" because their output is sent to a greater number of nodes, hubs also receive input from the same nodes, which co-determine their outputs. Consequently, information flows both from the bottom-up and from the top-down through the network. However, the more connected the hub, the more inputs it sums, and the less able it is to respond with unique outputs. While less well-connected nodes have a smaller "sphere of influence," their ability to distinguish between their inputs with unique outputs is significantly greater. Put another way, information flowing from the top-down is more "general" than information flowing from the bottom-up.

## MIDI Mapping

The MIDI Mapping layer allows users to map node output values to appropriate MIDI ranges. For example, if nodes are set to output 12 values:

1. output values from pitch nodes can be mapped to a chromatic scale (e.g. C4–C5);
2. velocity node outputs can be mapped to 10, 20, 30, 40 … 120 MIDI velocity values;
3. duration node outputs can be mapped to an arbitrarily chosen fixed range (e.g., 100, 150, 200 … 650 milliseconds) or a duration based on a subdivision of the entry delay times between notes.
4. entry delays values between notes are scaled to an appropriate musical range in milliseconds.

In addition to generating the basic attributes of notes, NetWorks provides for mapping network activity to MIDI cc control data to control various synthesis parameters such as filters, and so forth, chosen the user. Currently, however, these outputs do not feedback into the network.

Since NW MIDI data is computer-generated, sampled acoustic instruments are often used to give the music a "human feel" and help the listener compare the self-organizing output patterns to known genres. When mapping patterns to sound, and during mixing, the main goal is to preserve the integrity of the generated patterns by ensuring that the changing relationships between note attributes remain audible. Instruments with complex envelopes and textures, and effects (such as echo), were avoided.

## Evaluation of NetWorks Output

To date, two albums have been produced using the NetWorks system: "NetWorks 1: Could-be Music" and "Networks 2: Phase Portraits", which can be heard online:

- https://shawnbell.bandcamp.com/album/networks-1-could-be-music
- https://shawnbell.bandcamp.com/album/networks-2-phase-portraits

The most recent experiments can be found here:

- https://soundcloud.com/zomes

As mentioned, NW output dynamics range from complete order (and thus repetition without change) to complete chaos (and thus no element of predictability). The musicality of the output is greatest when the system is tuned to an intermediate between these extremes, i.e., the proverbial 'edge of chaos.' At this point there is a pleasing balance between familiar, repeating patterns, and novelty.

Shannon Entropy was also used to compare NW MIDI data sequences generated with rules having a random distribution of output values with MIDI data generated using LUTs that output (mostly) the same value when activated. Entropy was also used to compare NW pieces, tuned to the edge of chaos, to other genres of music to confirm subjective comparisons.

Entropy is a good measure of the unpredictability / complexity in data sequences. As a simplified data sequence, music has two features: the range of notes (pitch/duration pairs), and the repetitiveness of the notes. Entropy values capture the degree of variety and repetitiveness of note sequences in MIDI data. Roughly speaking, high entropy indicates surprising or unpredictable musical patterns while low entropy indicates predictable, repeating musical patterns (Ren 2015). In this analysis, the entropy of a piece was calculated by counting the frequency of musical events, specifically the appearances of each note (pitch-duration pair), as well as pitch and duration separately to get the discrete distribution of those events. Equation 1 was used to calculate the information content of each note. The expectation value of the information content, defined as $-logp(x_i)$, was used to obtain the entropy. The entropy is related to the frequency of musical events in a specific range. Differences in entropy values stem from differences of (1) the underlying possibility space size, i.e. how many different types of musical events there are, and (2) how repetitive they are. Although this does not take into account the order of events it provides a general characterization useful for comparing musical sequences (Ren 2015).

$$H(X) = -\sum_{i} p(x_i) logp(x_i), i \in n = outcomes \quad (1)$$

In Figure 2, the entropy value of ten NW pieces (x-tick=3) is compared with Bach's chorales (x-tick=1) and with jazz tunes (x-tick=2). In terms of entropy, NW pieces are closer to jazz than to Bach, which confirms informal subjective evaluations of NW music. X-tick=4 shows the entropy value for three NW pieces generated using a random distribution of LUT output values and x-tick=5 shows the entropy values of three NWs pieces with near uniform LUTs. These values verify the relationship between NW MIDI outputs and the LUTs that generate them.

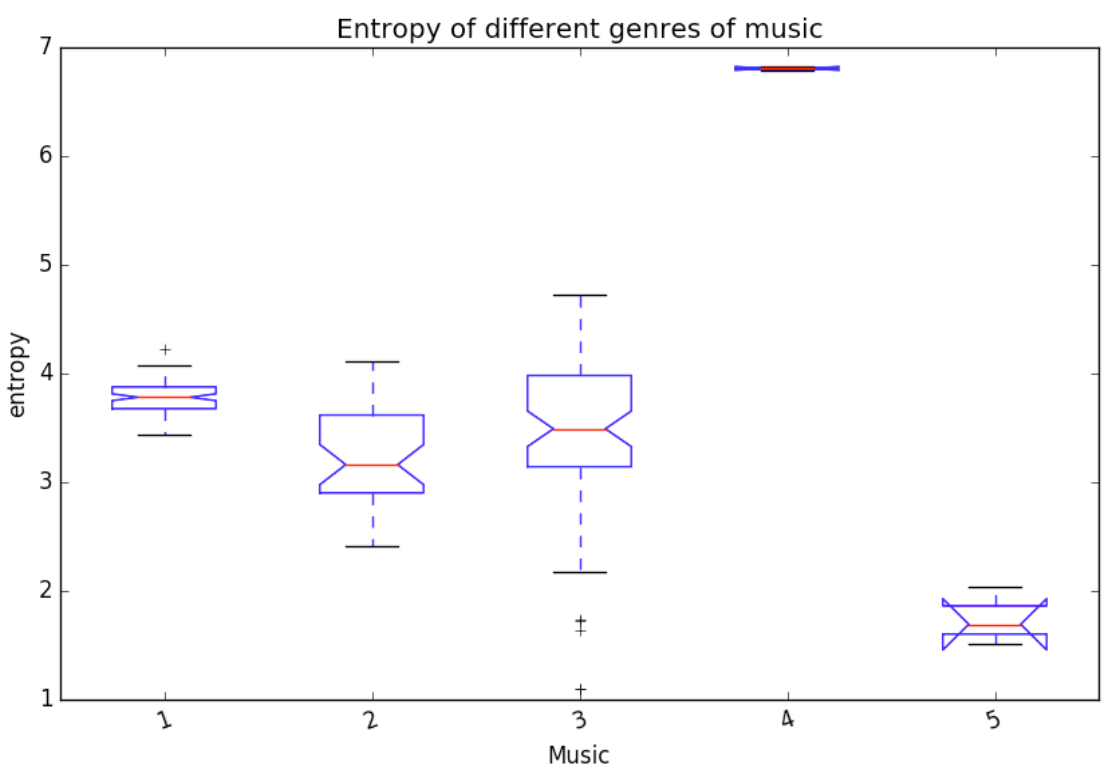

*Figure 2. Comparison of entropy of ten NW pieces (x-tick=3) with Bach chorales (x-tick=1) and jazz tunes (x-tick=2).*

Evaluation of NW music via social media (SoundCloud), shows an increasing interest in NW music from what is quite likely a diverse audience given the wide range of social-media groups to which NW music has been posted (e.g., classical, jazz, electronic, experimental, ambient, film music, algorithmic music, creative coding, complex systems, etc.). There has been a steady growth of "followers" over the two years (2014-2016) of posting NW pieces (28 tracks). As of the writing of this paper, NW has 323 followers. 7,893 listens, 831 downloads, 363 likes, 24 reposts, and 55 comments (all of which are positive).

As a search for "music-as-it-could-be," (e.g., new genres) a comment from SoundCloud indicates this goal may have been attained: "What can I say except I think I like it?" This suggests that the person has heard something they cannot categorize, but that sounds like good music.

## How NetWorks Implements Honing Theory

We now summarize how the NetWorks (NW) architecture and outputs adhere to and implement ideas from honing theory (HT), a theory of creativity inspired by chaos theory and the theory of complex adaptive systems (CAS).

### NW as Creative, Self-Organizing Structure

NW is hardwired to exhibit the key properties of real-world complex systems through its modular, scale-free, small-world properties. NW architecture has a shallow, fractal, self-similar structure (4 node, 16 node, and 64 node modules) that allows multiple basins of attraction to form in parallel, over different timescales, and interact.

NW networks are not neural networks; they do not adapt or learn by tuning weights between nodes through experience or training, nor do they evolve; nodes simply accept input and respond. Their LUTs do not change, adapt, or self-organize over time, but their dynamics do.

Just like an experience or realization can provide the 'seed incident' that stimulates creative honing, the pseudo-randomly generated initial conditions provide 'seed incidents' that initiate NW processing. After NW receives its inputs it is a closed system that adapts to itself (self-organizes). Musical ideas sometimes unfold in an open-ended manner, producing novelty and surprise, both considered hallmarks of emergence. A diversity of asynchronous interactions (sometimes spread out in time) can push NW dynamics across different basins of attraction. Idea refinement occurs when users (1) generate and evaluate network architectures, LUTs and mappings, and (2) orchestrate, mix, and master the most aesthetically pleasing instances of these outputs. The role of mental representation is played by notes—their basic attributes as well as attributes formed by their relationships to other notes.

**Cellular Automata-like Behavioral Classes**

NW nodes have a significantly different topology from Cellular Automata (CA). While CA have a regular lattice geometry, NW has a hierarchical (modular), scale-free, small-world structure. Moreover, unlike CAs, NW is updated asynchronously. However, similar to CA, NW exhibits Wolfram's class one (homogenous), class two (periodic), class three (chaotic), and class four (complex) behaviour (Wolfram 1984), and—rather than converging to a steady state—tends to oscillate between them. This is because the nested architecture of NW allow multiple basins of attraction to form in parallel and over different timescales. Pruning the scale-free architecture by reducing the inputs to hubs insulates clusters and modules from one another, reducing their interactions. Network dynamics within a basin of attraction can get pushed out of the basin by delayed values entering the system. In other words, because in the context of the current pattern an "old ideas" can push the dynamics to a different basin, the system exhibits "self-mending" behavior. This can result in musical transitions that lead to the emergence of new patterns and textures.

**Representational Redescription**

The network "makes sense" of its present in terms of its past by adapting to delayed values or "old ideas" entering the current pattern of activations. NW nodes hone by integrating and simplifying inputs from multiple sources, and returning a particular value. In NW, a catalyst or "catalytic value" is one that needs to be received on the inputs of one or more nodes to maintain one or more periodic structure (perhaps playing a different role in each). As NW strings notes together (often in parallel) in a stream of music, its nodes act on and react to both the nodes in their cluster, and to other clusters, via their hubs. Bottom-up and top-down feedback and time-delayed interactions are essential for an open-ended communal evolution of creative novelty.

Periodic structures are often disrupted (stopped or modified) by the introduction of a new (delayed) value, although sometimes this does not affect output. As interactions between nodes occur through entry delays, periodic musical structures unfold at different timescales. Slowly evolving periodic structures can be difficult to hear (due to intervening events) but can have a "guiding" effect on the output stream, i.e., they affect what Bimbot, Deruty, Sargent, and Vincent (2011) refer to as the "semiotic" or high-level structure of the music emerging from long term regularities and relationships between its successive parts. NW creates musical "ideas" that become the context for their further unfolding. Asynchrony, achieved by the (dynamically changing) values of the nodes in the Entry Delay Module allow previously calculated node values (including their own) to be output later in time. NW outputs both manifests the dynamics of the network, and in turn generate the dynamics. As with the autopoietic structure of a creative mind, NW is a complex system composed of mutually interdependent parts.

Let us examine how a NW network could be said to take on characteristics of an autocatalytically closed creative mind. The nodes collectively act as a memory in the following sense. When a node is activated, it sums the last values received on its inputs and uses the sum to output the stored value (which is then delayed before being sent to receiving nodes). Nodes are programmed so that their individual inputs can only store or "remember" the last value received. However, because nodes have 3, 4, 5, 14 and 39 inputs (excluding their own), and the network is asynchronous, a node (as a whole) can "remember" values spread out over time. How long a node can remember depends on its own ED value and the ED values of the nodes that participate in co-determining its output. It is important to note, however, that nodes can also "forget" much of the information they receive, if, for example, it receives different values on the same inputs since only the last ones are used when the node is activated. Again, how much it forgets depends on its own ED value and those of nodes to which it is linked. These memory patterns are distributed across the network. They are self-organizing because they can recur with variation, such that the whole is constantly revising itself. NW chains items together into a stream of related notes / note attributes. As NW strings notes together in a stream of music, its nodes are acting on and reacting to (feeding-back and feeding-forward information) to and from both the nodes in their cluster and to other clusters via their hubs. It would seem that bottom-up, top-down and time-based interaction / feedback are essential for an open-ended communal evolution of creative novelty.

### Contextual Focus and the Edge of Chaos

Some of NW's music sounds uninspired; it contains no surprising pattern development (e.g., a sudden transition or gradually modulated transition in texture, mood, or tempo), and/or the patterns do not elicit innovative variations. To minimize this problem, NW uses an architecture that, in its own way, implements contextual focus. Clusters of nodes that are more interlinked and share similar LUTs process in a more analytic mode. Hubs, which connect clusters into a small-world network and merge more distantly related musical ideas, process in a more associative mode. Because clusters have fewer inputs than hubs they are more discriminating than hubs. Hubs act as funnels, summarizing or simplifying the information they receive from multiple sources. Thus NW is hardwired to shift between analytic and associative modes by modulating the relative influence of top-down and bottom up processing.

NW structures transform as they propagate in time. As mentioned above, all four behavior classes have been observed. Class one and two dynamics do not change unless disrupted. When NW processes 'associatively' the output streams exhibit class two behaviour. When NW processes 'analytically' it exhibits Class three (deterministic chaos) behavior, which does not repeat if unbounded. Class four (edge of chaos) balances change and continuity.

Network dynamics often sound chaotic at the beginning of a piece–set in motion from an arbitrary, initial configuration ('seed incident'). Repetition and development of motivic materials and/or melodic lines then moves the system toward one or more attractor(s) (or "grooves"), resulting in a more stable, organized musical texture. Nodes with different rules of interaction are apt to disturb the system, pushing it into another basin. If it returns to a basin, a similar texture returns. When tuned to midway between order and chaos, the global stable dynamics are repeatedly disturbed. This pushes it either (1) into another basin, creating a transition to contrasting musical material, or (2) further from the attractor, to which it tries to return. NW exhibits something akin to goal seeking behaviour in how it moves toward or away from an attractor by keeping within a range of "desirable" values. This is similar to the use of functional tonality in western music, in which a piece departs and returns to its tonal center. Quasi-periodic dynamics provide a sense of organization through cycling musical textures, or a loose theme and variation structure. Disturbances may be caused by nodes with different rules of interaction, or by delayed values entering the stream. One factor that affects the aesthetic quality of the output is the mapping of the node output values to a specific ED scale (mapped values are used to delay node outputs). This appears to produce a balance between current events and older ones that is at the proverbial edge of chaos.

## Conclusions and Future Directions

NW's unique architecture—in particular, its scale-free network and transparent relationship between rules of interaction (LUTs) and MIDI output—was inspired by the science of complex adaptive systems as advocated by the honing theory of creativity. Its dynamics lie midway between order and chaos, and evolve—not through a Darwinian process (c.f., Gabora, 2005; Gabora & Kauffman, 2016)—but in the sense of generating cumulative, adaptive (in this case, aesthetically pleasing) change.

The number of possible LUTs that can generate 'edge of chaos' dynamics is extremely large. In the future we will expand the scope of NW to get a sense of to what extent the agreeable sound palette contributes to the aesthetic assessment. "By hand" rule design and "by ear" verification of the results will be augmented by evolutionary programming techniques guided by quantitative analyses. NW will also continue incorporating principles of HT. In turn, grounding the theory using NW is inspiring new developments in the understanding of creativity.

## Acknowledgements

Many thanks to Iris Yuping Ren for the entropy analysis. This research was supported by a grant from the Natural Sciences and Engineering Research Council of Canada.